\documentclass[12pt]{article}
\usepackage{amsmath}
\usepackage{amssymb}
\usepackage{amsthm}
\usepackage{amsfonts}
\usepackage[dvips]{graphicx}
\def\abstract#1{\vskip 7mm
        \begin{center}{\large Abstract}\par \smallskip
                \begin{minipage}[c]{12cm}
                        \small #1
                \end{minipage}
        \end{center}
}

\def\title#1{\begin{center}{\Large\bf #1}\end{center}}
\def\author#1{\vskip 5mm \begin{center}{#1}\end{center}}
\def\address#1{\begin{center}{\it #1}\end{center}}

\newcommand{\ssmatrix}[4]%
{\begin{pmatrix} #1 & #2 \\ #3 & #4 \end{pmatrix}}
\makeatletter
\@ifundefined{lesssim}{}{}
\@ifundefined{gtrsim}{}{}
\def\vereq#1#2{\lower3pt\vbox{\baselineskip1.5pt \lineskip1.5pt
\ialign{$\m@th#1\hfill##\hfil$\crcr#2\crcr\sim\crcr}}}
\makeatother

\begin{document}
\setlength{\baselineskip}{16pt}
\title{%
Topological derivation of Black Hole entropy \\
by analogy with a chain polymer
}
\author{%
  Masaru Siino\footnote{E-mail:msiino@th.phys.titech.ac.jp}
}
\address{%
  Department of Physics, Tokyo Institute of Technology, \\
  Oh-Okayama, Megro-ku, Tokyo 152-, Japan
}

\abstract{
The generic crease set of an event horizon possesses anisotropic structure though most of black holes are dynamically stable.
This fact suggests that a generic almost spherical black hole has a very crumpled crease set in a microscopic scale though the crease set is similar to a point-wise crease set in a macroscopic scale. In the present article, we count the number of such micro-states of an almost spherical black hole by analogy with an elastic chain polymer. This estimation of black hole entropy reproduces the well-known Bekenstein-Hawking entropy of a Schwarzschild black hole.}

\section{Introduction: A topological viewpoint of Event Horizon}
One of the most remarkable aspects of the Black Hole entropy\cite{BS} is that it is not proportional to something like a volume of a black hole but to the area of its event horizon, while entropy is an extensive variable in statistical mechanics.
Furthermore if one try to find appropriate volume like variable, even for a Schwarzschild event horizon, there is no such thing as the volume inside the horizon since it depends on the global solution one choose, which could even render the volume infinite, for a Cauchy surface. 

From one point of view, this is realized that the entropy is not of the black hole but of the event horizon which is the boundary of the black hole region.
In this sense, some authors have derived the black hole entropy by calculating the degrees of freedom only on the event horizon in various quantum theories, e.g., in quantum geometry\cite{QG} and by a technique in the setting of ADS/CFT correspondence\cite{AC}, and others discuss the statistical meaning of the boundary in the context of entanglement\cite{EE}.
Furthermore, the technique of Euclidean path integral\cite{GH} is also on this standpoint as the relevant contribution to the black hole entropy comes from boundary integration at the event horizon.

On the other hand, a recently developed technique of D-brane to derive the black hole entropy seems to be related to the whole of the black hole spacetime in itself\cite{SV}, though it is not fully clear what is estimated by this derivation. In this sense, it may be valid to regard the black hole entropy as the entropy of the black hole region, after all.

If the black hole entropy is really of the black hole region, we will need a reason why it is proportional to the area of the event horizon.
In this article, we try to estimate entropy of matter which have been absorbed into a black hole region during the black hole formation, relating it to the topological structure of its event horizon.
Then we would show that the black hole entropy is proportional to the area of the event horizon. Here we never relate the entropy directly to the area of the event horizon. The entropy concerns only a mass of the black hole.
Moreover, this entropy would be able to be regarded as the count of the ways to form a final black hole.


When we concentrate on the topological feature of an event horizon, which can be reduced to the structure of the crease set of the event horizon\cite{MS1}. 
On the crease set, two or more generators of the event horizon intersect and the event horizon is not smooth\cite{BK} (the rigorous definition will be given in the third section).
Furthermore, the catastrophe theory\cite{SK}\cite{MS2}\cite{AN} tells that the generic crease set is composed of not a point-wise structure of a spherical black hole but two-dimensional structures and their bifurcations. Taking an appropriate timeslice, this two-dimensional crease set provides a toroidal event horizon.
In this sense, the spherical topology of event horizon is structurally unstable. 

Here it should be noted that the above observations do not mean that a black hole and its crease set are always highly anisotropic. Since the catastrophe theory suggests
 that the 
spherical topology changes under
small perturbations in a corresponding microscopic scale, the degree of anisotropicity
would be very small in some cases.
For example, when almost spherically symmetric matter collapses to an
 almost
spherical black hole, in microscopic scale its crease set will be highly distorted and
bifurcated and its event horizon will have very complicated topology.
On the contrary, in a macroscopic scale, the crease set can be treated approximately as a point and then the event horizon seems to have a spherical topology.

These aspects make us expect that the crease set will be endowed with  micro-canonical entropy.
In the present article, we estimate the entropy of the crease set by
analogy with a chain polymer, since the one-dimensional crease
set possesses similar structure to the chain polymer (and two-dimensional one
will be similar in micro-canonical statistics).
Assuming that a multiply folded crease set forms zigzags like the
chain polymer, we can estimate the micro-canonical entropy of the crease set.
Then we achieve entropy of almost spherical black hole, which is coincident with the Bekenstein-Hawking entropy. Finally we are going to interpret this entropy as the missing information of falling
matters. 
In other words, the entropy counts the ways to form a final black hole.

In the next section, we recall the way to calculate the
entropy of a chain polymer in a simple Ising model.
The third section shows how one can estimate the
entropy associated with the black hole from the view
point of its topological structure.
The final section is devoted to summary, discussions and speculations.

\section{Entropy of Chain Polymer}
\begin{figure}
\begin{center}
\includegraphics[width=11cm,clip]{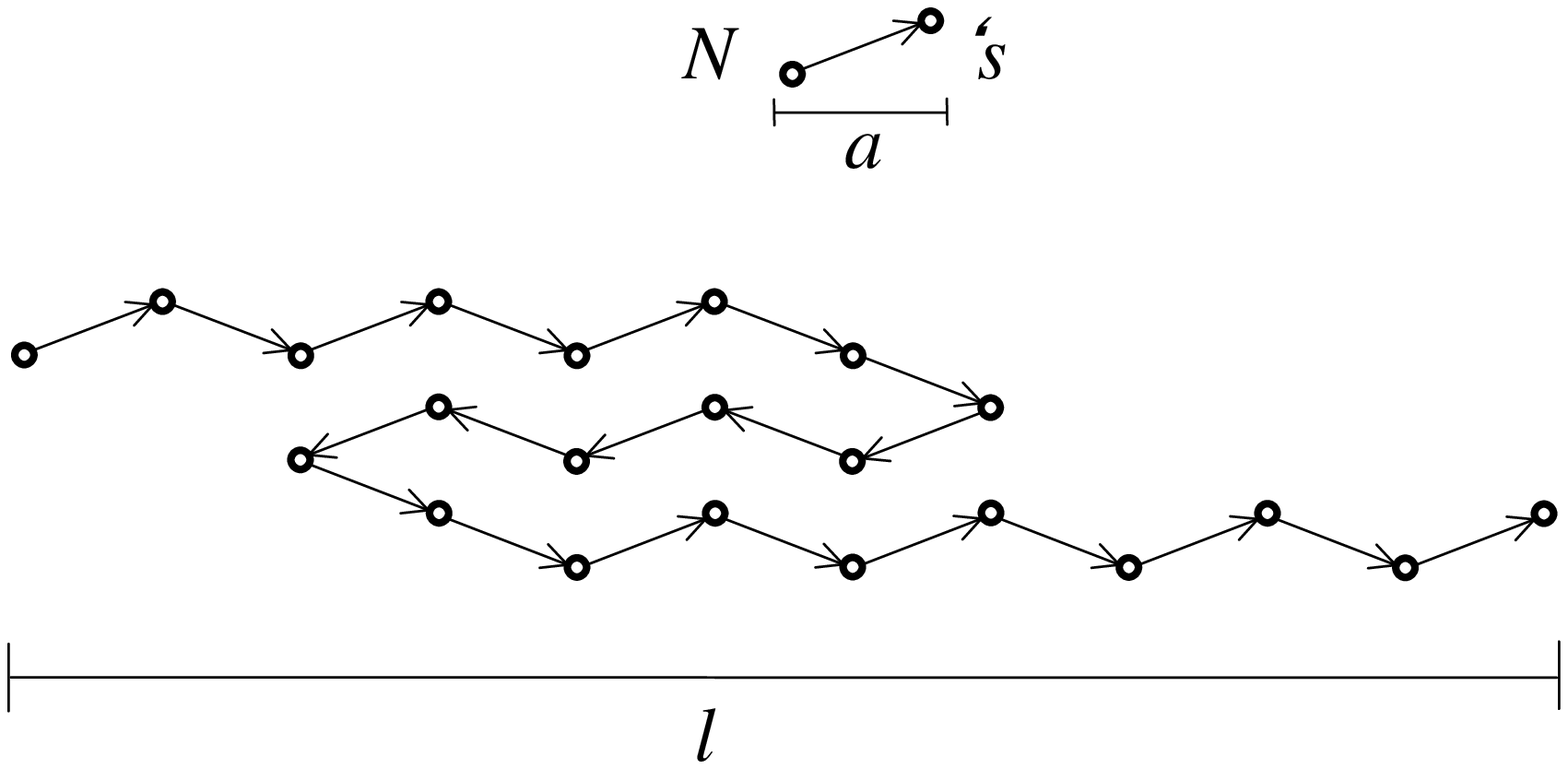} 
\end{center}
\caption{A simple Ising model of the elasticity of a chain polymer is 
exemplified. A chain polymer is composed of $N$ monomers with a length $a$.
Each monomer can be directed only to the left or the right.
Though the maximum length of the chain should be $Na$, the highly folded
chain is with a length $l\ll Na$.}
\label{fig:pm}
\end{figure}

In this section, we recall a simple Ising model of elasticity of a chain polymer\cite{IM}.
Suppose a large number $N$ of monomers with a length $a$ form a chain polymer with
 a total length $Na$. Furthermore, suppose this polymer is folded into an arbitrary length $l\ll Na$.
To give a simple model of folding, we suppose that 
each element of the monomers can be directed only to the right or the left 
in equal probabilities as exemplified in figure \ref{fig:pm}.

In micro-canonical statistics, the length $l$ is a parameter describing a state of the system.
The number of allowed configurations $W(l)$ is given by
\begin{equation}
W={2N! \over N_{\rightarrow}!N_{\leftarrow}!}={2N!\over(\frac12N-l/a)!
(\frac12N+l/a)!},
\end{equation}
where $N_{\rightarrow}$ and $N_{\leftarrow}$ are the number of right- and left-directed elements, respectively.
Then, using the Stirling formula $\log N!\simeq N\log N -N$, the entropy of this polymer 
becomes
\begin{equation}
S=\log W \simeq N\log N -(\frac12N-l/a)\log(\frac12N-l/a)-(\frac12N+l/a)\log(\frac12N+l/a).
\end{equation}
Under the assumption that the length $l$ is much smaller than $Na$, 
this is approximated as 
\begin{eqnarray}
S(l)&=&N\log 2-\frac{l^2}{2Na^2}+O(N\cdot(l/Na)^4)\\
&=&S(l=0)-\frac{l^2}{2Na^2}+O(N\cdot(l/Na)^4),
\label{eqn:scp}
\end{eqnarray}
where we have used $\log(1+x)=x-x^2/2+... (x\sim l/Na\ll 1)$.

This gives a simple model of elasticity. Indeed, from the first law of thermodynamics
 $TdS=dU-fdl$ ($T$ and $U$ are the temperature and the internal energy), the elastic force $f$
obeys well known Hooke's law in the leading order:
\begin{equation}
f=T\left(\frac{\partial S}{\partial l}\right)_U=-\frac{Tl}{Na^2}\ .
\label{eqn:ef}
\end{equation}
Namely, the force is proportional to the temperature $T$. Actually, a rubber band 
contracts when it
is wormed up, while an iron wire expands.

\section{Black Hole Entropy}
Now we estimate entropy associated with the crease set of an event horizon.
Here we should give the definition of the crease set\cite{BK}.
We consider a null vector field $K$ on the event horizon which is tangent to 
the null geodesics  generator $\lambda$ of the event horizon.
$K$ is not affinely parameterized, but parameterized so as to be continuous
even on the endpoint where the caustic of $\lambda$ appears. Then the endpoints of $\lambda$ are the zeros of
$K$, which can become only past endpoints, since $\lambda$ must reach to infinity in the future direction. Of course, using an affine
parameterization, $K$ becomes ill defined at a subset of the set of the endpoints. We call such a subset the {\it crease set}. To be
precise, we define the crease set by the set of the endpoints contained by two or more null generators of the event horizon. Thus the set of the endpoints 
consists
of the crease set and endpoints contained by one null generator. 
The
closure of the crease set contains the set of the endpoints of 
the event horizon
generators, and the event horizon is indifferentiable there\cite{MS1}\cite{BK}.

From Ref.\cite{MS1}, the spatial topology of an event horizon in a timeslicing is determined only
by the timeslicing of the crease set. This implies that the crease set possesses all of the topological information of the black hole.
In other words, an event horizon is completely determined once we know the crease set and all of the light rays starting from the crease set, since the event horizon
should be their envelope.
Hence we expect that the crease set will give all of the global information about the event horizon, while the light rays can be determined only by a local geometry.
In this section, we try to estimate the entropy associated with that global
information carried by the crease set of the event horizon.

In our point of view, the entropy of the crease set is brought by the missing information of falling bodies when they fall beyond the event horizon of a black hole. Since the crease set is the multiple
point of the generator of the event horizon, its own structure will be changed provided that
a falling body affects a congruence of the generators of the event horizon when the falling body
crosses it. Then both of the topology of the event horizon and the structure of the crease set 
reflect some information included in the configuration of matter outside of the black hole.
If we suppose that the topology of a black hole finally settles to a single spherical 
one after all outside matter has fallen into the black hole, this information of the
topology of the event horizon turns out to be absorbed into the black hole and translated into the information of the crease set.
Therefore we expect that the missing part of this crease set information will correspond to the black hole entropy and try to estimate the entropy of the crease set.

To consider the degeneracy of the crease set,
one might assign the degrees of freedom to each Planckian scale segment of the
crease set as
the simplest model.
 The crease
 set, however, can be two-, one- or zero-dimensional. Each fundamental
 element becomes Planckian area or length or vanishes, respectively.
For example, the entropy of a one-dimensional crease set with a length $L$
will intuitively estimated as $
S=\log W=
\log (C^{L/l_{pl}})$, where $C$ is the number of possible states for each fundamental element. This is not what we have expected since 
$L$ could not be proportional to the area of the event horizon in the case of the one-dimensional crease set. 

On the contrary, by the analogy with a chain polymer,
we will derive entropy of the crease set proportional to the area of the event horizon in the following.
To determine entropy, we count the logarithm of the micro-state degeneracy. Though there may be various model of the micro-state, in the present article we apply following very simple Ising model, similarly to the chain polymer.

\begin{figure}
\begin{center}
\includegraphics[width=11cm,clip]{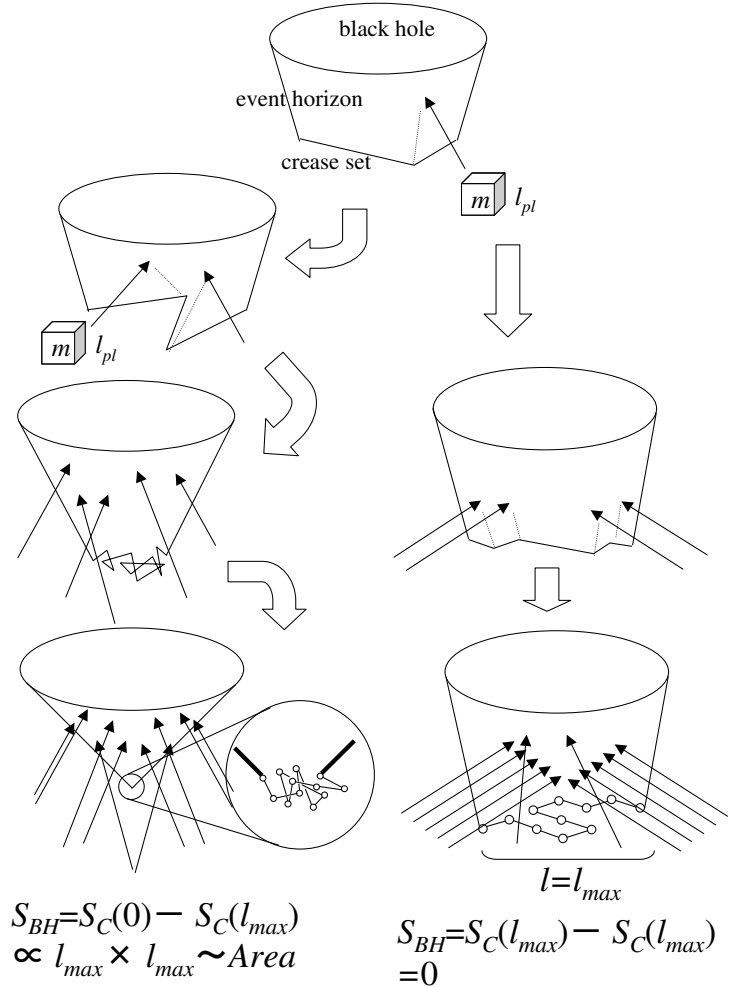} 
\end{center}
\caption{Two types of black hole formation are illustrated. The left branch is an almost spherically symmetric collapse.
A small volume element $l_{pl}^3$ with a mass $m$ affects a generator of the event horizon when it fall into the black hole. As illustrated on the top, this effect cause the bending of the
crease set of the event horizon. If many small bodies fall into the black
hole from random directions at
random time, the crease set is bent many times in various directions and
confined in a small region. The resultant black hole seems almost spherically  symmetric in macroscopic scale.
On the other hand,
the right branch is an extremely anisotropic collapse.
Since many of the falling bodies are ordered to be from a special 
direction, the bending effect will not make the crease set so small.}
\label{fig:ov}
\end{figure}

First, we consider only the case of one-dimensional crease set for simplicity
since the case of two-dimensional crease set will be different only by a factor in their entropy.
On the other hand, it is concluded that the point-wise crease set is not generic from the catastrophe theory \cite{SK}\cite{MS2}\cite{AN}. This implies that even for an almost spherically symmetric collapsing of matter,
 the matter and spacetime are not rigorously spherically
symmetric `in a microscopic scale' because of an anisotropic small perturbation.
This will cause a highly folded crease set, which is confined within a very small region. 
Then it is not point-wise in a microscopic scale but in a macroscopic scale (see the bottom-left of figure \ref{fig:ov}). 

There should be many ways to fold and confine the crease set. 
Considering a number of ideal small fundamental elements of
 the crease set to fold, 
this situation is very similar
to the chain polymer discussed in the previous section (compare figure \ref{fig:pm} and figure \ref{fig:ov}). Then we will count
the number of their allowed configurations and estimate its entropy, by the analogy with the
chain polymer. 

In the case of the chain polymer, its entropy $S_{CP}$ is given by (\ref{eqn:scp}) and 
we think that the entropy of the crease set $S_C$ is same as  $S_{CP}$;
\begin{equation}
S_C(l)=S_{CP}(l)=S_{CP}(0)-\frac{l^2}{2Na^2},
\end{equation}
where $l$ is the length of the crease set. $N$ and $a$ are the number and 
length of the ideal fundamental element, respectively.

In our discussion, the state with $l=0$ (the left branch of figure \ref{fig:ov})
 is regarded as an almost spherically
 symmetric black hole, since this state is macroscopically similar to a spherical black hole
 with a zero-dimensional (point-wise) crease set.
On the other hand, to make the black hole most anisotropic, collapsing matter
 must
 be most tilted in a special direction. This configuration will
not allow any degeneration of the micro-state. Nevertheless, the black hole would not be
allowed to take such an arbitrary tilted configuration; rather, 
it is natural that
there is an upper bound $l_{max}$ for $l$ since a black hole with infinitely large $l$ seems to be unphysical. Then if we have an upper bound
 $l_{max}$ (the right branch of figure \ref{fig:ov}), it will be valid to regard $S(l_{max})$ as the zero-point of the entropy of the black hole. Therefore the entropy of an almost spherical black hole is given by
 \begin{equation}
S_{BH}(l=0)\equiv S_{C}(l=0)-S_{C}(l_{max})=\frac{l_{max}^2}{2Na^2}\ .
\label{eqn:BHE}
\end{equation}

We may expect that the upper bound $l_{max}$ is about a final black hole 
mass $M$, since it is the only reasonable scale in a gravitational dust collapse. Furthermore, the hoop conjecture\cite{TH} requires the length of the crease set should be bounded by $2\pi M$\cite{ID}. Hence we assume $l_{max}\simeq 2\pi 
M$.
In addition to it, we should assume $l_{max}/a\ll N$ in order to derive eq.(\ref{eqn:scp}) in the previous section. The consistency and validity of 
this condition will be discussed later.
Consequently we observe that this entropy proportional to the area of the event
horizon ${\cal{A}}\propto M^2$.

By the way, eq.(\ref{eqn:BHE}) has an unfavorable factor $a^2/N$.
Expecting that this entropy coincides to the Bekenstein-Hawking entropy, $Na^2$ should be on a scale of $l_{pl}^2$.
This turns out that $a$ is $\frac{l_{pl}}{\sqrt{N}}\ll l_{pl}$ as $N$ is a very large number.
Since it is unreasonable to give a much smaller structure than Planckian length to quantum spacetime, we cannot accept such a small $a$.

\begin{figure}
\begin{center}
\includegraphics[width=11cm,clip]{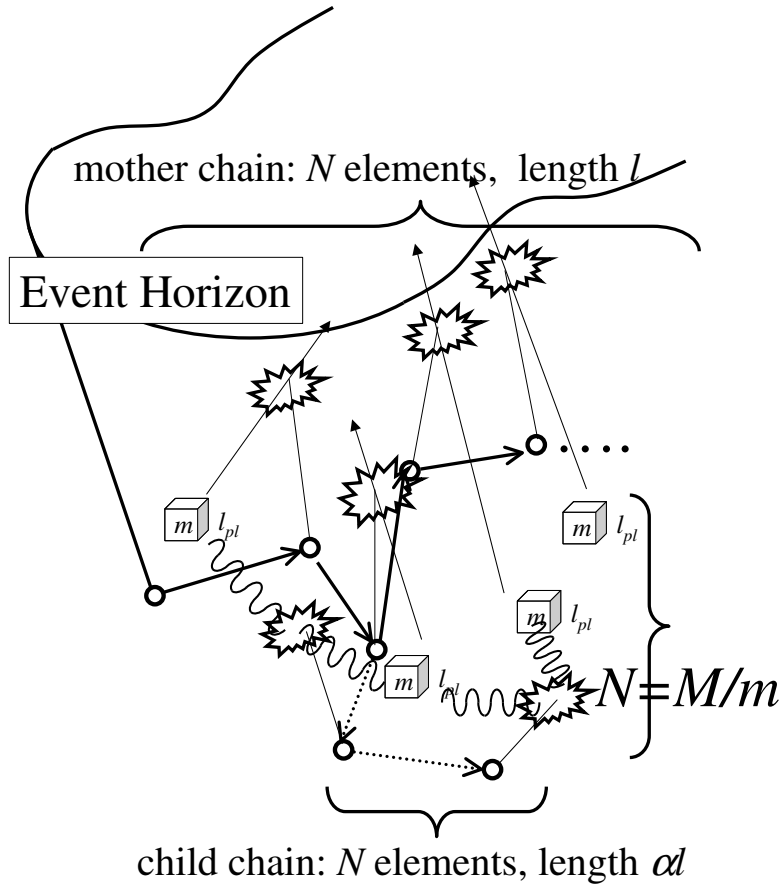} 
\end{center}
\caption{The number of the hinges of the mother chain is $N$, since
the number of falling volume elements is $N=M({\rm total\ mass})/m({\rm a\ mass\ of
\ a\  volume})$. Bold (dotted) arrows are the segments of the crease set forming the mother (child) chain. A child chain comes up from the hinge of the mother chain in a probability $\beta$.
Each small volume element of matter fall into the black hole along arrows
directed above.
They cross the event horizon and affect its generators (narrow vertical line) at exploding
symbols. On the other hand, three point interactions affect the generators along waving lines. The child chain forms $N$ hinges by these three point interactions.}
\label{fig:mb}
\end{figure}

This problem of the scale of a small segment is resolved by considering the branches of the crease set.
As pointed out in \cite{SK}\cite{AN}, there are possibilities that the crease 
set is branched at a hinge where the crease set can angle.
We assume that a new branch (child chain) with a length $\alpha l$ ($\alpha$ is less than one,  since a child should be smaller than its mother by definition) comes up at 
some hinges in a probability $\beta$ (see also figure \ref{fig:mb}) and is composed of $N$ elements; this will be justified later. The number of such a branch
is given by the probability and the number of mother's hinges as $\beta N$. 

Moreover there are also grandchildren and further descendants.
Naively, the number of $n$-th descendants might be considered to be 
$(\beta N)^n$ in geometrical progression.
However this is not realistic because we will require infinite volume to embed
 all the family of geometrical progression $\Sigma_i^{\infty} (\beta N)^i$.
As suggested later, it seems that this divergence is related to the divergence
of many-body interaction. Then we would require any regularization for this divergence. Since the territory of a child and its all descendants would be limited
around each of $N$ hinges of the mother chain,
we assume that the number of $n$-th descendants is $\beta^nN$ rather
than $(\beta N)^n$ as a regularization. By this assumption, the total length of the family becomes
$N\Sigma_i^{\infty} \alpha^i\beta^i l$ and will converge so as to be embedded, 
since the $n$-th descendant is with
a length $\alpha^n l$.
  Then the total number of degenerated micro-states is given by
\begin{equation}
W_{tot}=W(l)\cdot W(\alpha l)^{N\beta}\cdot W(\alpha^2 l)^{N\beta^2}...,
\label{eqn:wto}
\end{equation}
where $n$-th factor is the contribution of all ($n-1$)-th descendants.
The entropy of a crease set (\ref{eqn:BHE}) is changed by a factor and now we are not worried about the factor $1/N$ any more: the total entropy is
\begin{eqnarray*}
S_{Ctot} = \log W_{tot} &=& S_0-\frac{l^2}{2Na^2}-\frac{(\alpha l)^2}{2Na^2} 
N\beta -\frac{(\alpha^2 l)^2}{2Na^2} N\beta^2-...\ +O(l^4)\\
&=& S_0-\frac{l^2}{2a^2}(\frac1{N}+\alpha^2\beta+\alpha^4\beta^2+...+\alpha^{2n}\beta^{n}+...)+O(l^4)\\
& \sim & S_0- \frac{\alpha^2\beta}{1-\alpha^2\beta}\frac{l^2}{2a^2},\\
S_{BH} &=& \frac{\alpha^2\beta}{1-\alpha^2\beta}\frac{l_{max}^2}{2a^2},
\end{eqnarray*}
where $S_0$ is the sum of all $l$-independent terms.
On the third line, it is supposed that $N$ is sufficiently large. Though the infinite sum might have
any cut off, it would change the result only by a numerical factor of the order of magnitude of one.

If we rigorously require $S={\cal{A}}/4l_{pl}^2=\pi M^2/l_{pl}^2$,
the relation
\begin{eqnarray*}
\frac{4\pi^2M^2}{2l_{pl}^2}\frac{\alpha^2\beta}{1-\alpha^2\beta}&=&\frac{\pi
 M^2}{l_{pl}^2}\\
\frac{\alpha^2\beta}{1-\alpha^2\beta}&=&\frac1{2\pi},
\end{eqnarray*}
will determine $\alpha^2\beta$,
since hoop conjecture says $l_{max}\sim 2\pi M $ and $a$ should naturally
 be $l_{pl}$.

Now we must discuss the case of a non-chain-like crease set.
Indeed, Refs \cite{MS2}\cite{SK}\cite{AN}\cite{W} tell us that it is important to
 consider a crease set with
two dimensions. The discussion of a two-dimensional crease set can be proceeded
with, as following.
Intuitively, the two-dimensional crease set has two independent degrees of freedom to fold.
This will make the state counting the square of that of a one-dimensional
 crease set and its entropy two times.
For further accurate estimation, it might be valid to discuss in the theory of random surface. Similarly, in the case of the random surface, the regular term
 of its entropy around $l=0$ is also proportional to $l^2$\cite{ZA}.
Then the elastic force is always proportional to the amount of its deformation (see eq.(\ref{eqn:ef})) independently of its form, size and dimensions.
This is consistent with the general Hooke's law, i.e. a stress tensor is proportional to a distortion tensor. This consistency makes us convinced that our
estimation is valid independently of the form, size and dimensions.

So we summary the estimation as
\begin{equation}
S_{BH}=F(n)G(\alpha^2\beta)\frac{\cal{A}}{4l_{pl}^2},
\end{equation}
where $F(n)$ and $G(\alpha^2\beta)$ are numerical factors of the order of magnitude of  one determined by
 the dimensions and branching of the crease set, respectively.

Finally, we discuss the assumptions we have made above.
Here we should discuss the meaning of $N$ and the validity of the assumptions about the amount of it. In the present estimation we have supposed that the number of mother's elements $N$ is a fixed large number, and $0<l/a<l_{max}/a$ is
 much less than $N$. 

One may be doubtful that these assumptions are consistent to physical situation. 
To make clear this point, we consider the relation between $N$ and the falling
bodies as following and illustrated in figure \ref{fig:ov}. We think an ideal process in which some small elements with a volume $l_{pl}^3$ and a mass $m$ fall into a black hole.  
The top of figure \ref{fig:ov} illustrates that a falling body gravitationally deforms a generator of the event horizon, and consequently the crease set will form a hinge and be angled there.  Here we note that the formation of the hinge
occurs before the falling of the body in the sense of usual spatial timeslicing. 
Since the event horizon, however, is defined as the boundary of a past set,
the mass of the falling body affects a past part of the event horizon along its null generators.

If many bodies randomly fall into the black hole (see the left branch of figure
\ref{fig:ov}), the crease set will be repeatedly angled in various directions and finally confined into a small region.
Therefore we guess that almost spherical collapse can occur through such a random falling process of a large number of 
small bodies. On the other hand, if the
small bodies are not random but ordered to be anisotropic in a special 
direction (the right branch of figure \ref{fig:ov}), the crease set
will be more spread and the anisotropic black hole is formed. Hence the 
entropy of the crease set is related to
that randomness of falling bodies.
To determine $N$, it will be valid to relate the number of
the hinges of the crease set and the falling ideal volume elements 
with a volume $l_{pl}^3$,
into which the collapsing matter could be decomposed.

Simply, we regard the number of the collapsing ideal volume 
elements as the number of the hinges of the mother chain $N$.
The consistent interpretation of the child and descendant chain is 
following (and see figure \ref{fig:mb}).
When a falling body crosses event horizon generators, the mother chain is angled by the falling body directly. There comes up
a child chain (dotted arrows) in a probability $\beta$.
Besides, this child chain is also affected by another falling volume 
element
through three-point interaction (among one event horizon generator, one 
body making the child chain, and another body) since gravitation is long range force. Therefore we consider that the hinges of the child chain are formed by this three point interaction. It is well known that such many-body interactions diverge and
need any regularization. Here we think that this regularization corresponds to
the assumption that the number of $n$-th descendants is $\beta^n N$ rather
than $(\beta N)^n$.
Then the hinges of the child chain are assigned the two falling volume elements (indicated by waving lines in Fig.\ref{fig:mb}); one of them has made the child chain. Hence the child chain possesses $N$ hinges. Similarly, 
a $n$-th descendant chain also forms about $N$ hinges under the influences of $(n+1)$ different falling volume elements.
These pictures give an explanation to the formation and number of the hinges of the descendant chain.  

Now we consider the number of elements $N\sim M/m$, inheriting from the number of falling volume elements.
The mass $m$ of the volume element of $l_{pl}^3$ should be much smaller than Planckian mass so that it will not be a black hole but an ordinary matter.
Then we have following inequalities,
\begin{equation}
N\sim \frac{M}{m}\gg\frac{M}{l_{pl}}\sim\frac{l_{max}}{l_{pl}=a}>\frac{l}{a}.
\label{eqn:NJ}
\end{equation}
Therefore we have confirmed all parameters are in a realistic range and
the assumptions are consistent. 

The picture illustrated in figure \ref{fig:ov} might be something 
kinematical while the process we think of is dynamical.
In other words, the picture put an interpretation that this black hole entropy counts the logarithm of the number of the ways to form an almost spherical black hole.
\section{Summary, Discussions and Speculations}
In the present article, we have argued that the Bekenstein-Hawking entropy
of the Schwarzschild black hole can be derived independently of the area of the event horizon as
 the entropy of its
 crease set.
This put an interpretation on the black hole entropy, i.e. it measures the missing topological (global) information
of the collapsing matter corresponding to the configuration of the falling volume elements in spacetime.

We have considered only Schwarzschild black hole as a final state of gravitational collapse. One may feel that it is important to extending the result  to a rotating black hole.
At present, however, we cannot imagine what shape of an event horizon is appropriate to compare with the Kerr black hole, as a zero of the entropy.
A  discussion of a chain polymer also should be changed.
To discuss these problems, we should relate the angular momentum to any character of the crease set,
as the mass of black hole have been related to the maximum length $l_{max}$
 of the crease set.

Moreover we would like to comment on the origin of the entropy estimated in the present article.
As discussed in the end of the previous section, the entropy is related to the falling bodies.
To be concrete, the entropy measures the disorder of the position and velocity of the falling bodies. Of course, these are not all of the information that falling bodies carry. In other words, the black hole entropy could be directly related to only the entropy of this disorder. The black hole entropy is the logarithm of the number of the possible configurations of falling matter to 
form a final Schwarzschild black hole, if we can decompose the falling matter 
into ideal small volume elements $l_{pl}^3$ with mass $m\ll m_{pl}$ and omit the process that tilted black holes settle to a Schwarzschild black hole 
by radiating gravitational waves.

Finally, we estimate the upper bound of a thermal elastic force of the crease set.
Substituting the Hawking temperature $T_H\sim1/M$ into $f=-T\partial S/\partial l(l=l_{max})$, we observe $f\sim1/a^2$ is independent of $M$.
Though its realistic meaning is not clear, this aspect coincides with the fact that the failure of Hooke's law occurs independently of the scale or form of the elastic body. 
Here we speculate that this coincidence implies the validity of the present
discussions (especially, the assumption $l_{max}\sim 2\pi M$ from the hoop conjecture).
The mechanism of the failure of a black hole formation or the naked singularity formation, which is the ground of the hoop conjecture, might be realized by analogy with the existence of such an elastic limit.

As a reader is noticed, the present estimation does
not work in different spacetime dimensions. This is because of the absence of the hoop conjecture in the other
spacetime dimensions. In turn, that fact might be able to speculate new conjectures in other spacetime dimensions if we require that this estimation reproduce
the Bekenstein-Hawking entropy also in other spacetime dimensions.

\section*{Acknowledgments}
This work is based on another research with Dr. Koike.
The author thanks to Prof. R. M. Wald for his helpful discussion.

\flushright

\end{document}